\documentstyle[stwol,psfig,epsf]{article}
\epsfverbosetrue

\def\simgt{\stackrel{>}{{}_\sim}}

\def\be{\begin{equation}}
\def\ee{\end{equation}}
\def\bea{\begin{eqnarray}}
\def\eea{\end{eqnarray}}

\begin{document}

\title{ELECTROWEAK INTERACTIONS -- THEORY}

\author{S. POKORSKI}

\address{Institute of Theoretical Physics,
Warsaw University, 00-681 Ho\.za 69, Warsaw, Poland}

\twocolumn[\maketitle\abstracts{Plenary talk given at the 
XXVIII International Conference
on High Energy Physics, 25-31 July, 1996, Warsaw, Poland.}]
 
In the first part of this talk 
there is given a very brief review of
the status of the Standard Model (SM). In the second part I
discuss electroweak interactions and physics beyond the SM.

\section{Status of the Standard Model}

The status of the SM is measured by a  comparison of precision 
data with precision calculations. The data are summarized by Alain
Blondel \cite{BLON} and the precision calculations have been reviewed in
the parallel session \cite{HOL}. Not only complete one-loop
calculations of the electroweak observables are available
(already for several years) but also a steady progress is being
made in including the dominant higher order effects (QED
corrections, QCD corrections, top quark mass effects,
resummation of the leading terms, (some) 2-loop weak
corrections, $\dots$). I'll limit my discussion of those matter
to a few remarks on selected topics, of particular interest
during this conference.

\subsection{Muon anomalous magnetic moment}

\indent One of the best measured electroweak observables is the 
muon anomalous magnetic moment
\begin{eqnarray}
a_\mu =(g_\mu -2)/2=(116 592 300 \pm 850)\times 10^{-11}\nonumber
\end{eqnarray}
The present accuracy, $\sim 10^{-3}\%$, is by factor two better
than the accuracy of the results for the Fermi constant $G_\mu$
and the $Z^0$ boson mass $M_Z$. Still, this accuracy is sufficient
to test only the QED sector of the SM \cite{SEE}. A renewed interest
in the muon anomalous magnetic moment is due to the forthcoming
Brookhaven National Laboratory experiment, with the
anticipated accuracy $\pm40\times 10^{-11}$. A test of the
one-loop weak corrections $\sim 195\times 10^{-11}$ becomes then, in
principle, possible. The weak 2-loop terms have recently been
also calculated \cite{CZAR} and found to be small, giving $(152\pm
3)\times 10^{-11}$ for the combined one- and two-loop correction.
Unfortunately, the theoretical precision of the SM prediction is
overshadowed by the large uncertainty in the hadronic photon
vacuum polarization \cite{EID}, $\pm 153\times10^{-11}$, (i.e. of 
exactly similar magnitude as the weak corrections and by factor 3 
larger than the 
expected experimental error!) and (to a lesser extent) by 
some controversy on the contribution from the light-by-light
scattering mediated by quarks, as part of the 3-loop hadronic
contribution \cite{HAY}. In conclusion the present comparison between 
the theory and experiment, $a^{exp}_\mu - a^{th}_\mu =(543\pm
865)\times 10^{-11}$, is dominated by the experimental error. With
the anticipated accuracy, $\delta a^{exp}_\mu = \pm 40\times
10^{-11}$, it will be dominated by the uncertainty in the
hadronic vacuum polarization, and the weak corrections can be
tested only after a substantial reduction of this uncertainty.
This can only be achieved by new measurements of the cross
section for $e^+e^-\rightarrow hadrons$ in the low energy range.
Under the same condition, the precise measurement of $a_\mu$ will be
a very important test of new physics, sensitive to mass scales
beyond the reach of the present accelerators \cite{NATH}. (Even the
present result puts some constraints, though mariginal, on the
parameter space in the Minimal Supersymmetric Standard Model.)

\subsection{$Z^0$-pole observables}

Turning now to the bulk of the electroweak precision
measurements ($M_W$, $Z^0$-pole observables, $\nu e$, $lp$, $\dots$)
one should stress that the global comparison of the SM predictions with
the data shows impressive agreement. Both, the experiment and the theory
have at present similar accuracy, typically 
${\cal O}(1$ $^{_{0}}\!\!\slash\!_{_{00}}$)! The
dramatic change in the data, reported at this conference, are the
new values for $R_b$ and $R_c$ discussed in detail in the
previous talk. Both are now in agreement with the SM.  The
predictions of the SM are usually given in terms of the
very precisely known parameters $G_\mu$, $\alpha_{EM}$, $M_Z$
and the other three parameters $\alpha_s(M_Z)$, $m_t$, $M_h$.
The top quark mass and the strong coupling constant are now also
reported from independent experiments with considerable
precision:
$m_t=(175\pm 6)$ GeV and $\alpha_s(M_Z)=0.118\pm 0.003$, but those
measurements are difficult and it is safer to take $\alpha_s,
m_t, M_h$ as parameters of an overall fit.
Such fits give values of $m_t$ and $\alpha_s$ very well consistent
with the above values \cite{BLON}.

The theoretical uncertainties in the SM predictions (for fixed 
$m_t$, $M_h$, $\alpha_s$) come mainly from the RG evolution of
$\alpha_{EM}\equiv\alpha(0)\rightarrow\alpha(M_Z)$
(to the scale $M_Z$) which depends (again!) on the hadronic
contribution to the photon vacuum polarization $\alpha(s)
=\alpha (0)/(1-\Delta\alpha(s))$ where
$\Delta\alpha(s)=\Delta\alpha_{hadr}+\dots$ and
$\Delta\alpha_{hadr}=0.0280\pm 0.0007$ \cite{EID} (unfortunately, 
it is a different function of $\sigma(e^+e^-\rightarrow hadrons)$ 
than  for the muon anomalous magnetic moment). The present
error in the hadronic vacuum polarization propagates as 
${\cal O}(1$ $^{_{0}}\!\!\slash\!_{_{00}}$)
errors in the final predictions.
The other uncertainties come from the neglected higher order
corrections and manifest themselves as renormalization scheme
dependence, higher order arbitrariness in resummation formulae
etc. Those effects have been estimated to be smaller than
${\cal O}(1$ $^{_{0}}\!\!\slash\!_{_{00}}$), hence the conclusion 
is that the theory and experiment agree with each other at the 
level of ${\cal O}(1$ $^{_{0}}\!\!\slash\!_{_{00}}$) accuracy. In
particular, the genuine weak loop corrections are now tested at
${\cal O}(5\sigma )$ level and the precision is already high 
enough to see some sensitivity to the Higgs boson mass.

\subsection{The Higgs boson in the SM}

The electroweak observables depend only logaritmically 
on the Higgs boson mass (whereas the dependence on the top 
quark mass is quadratic). Global fits to the present data give
$M_h\approx145^{+160}_{-80}$ GeV and the 95$\%$ C.L. upper bound is
around 600 GeV \cite{ELLIS}. Thus, the data give some indication for 
a light Higgs boson. (It is worth noting that $M_h=1$ TeV is 
$\simgt3\sigma$
away from the best fit). The direct experimental lower limit on $M_h$ is
$\sim 65$ GeV. These results should be placed in the context of the 
theoretical lower and upper bounds for the SM Higgs boson mass
(for extensive list of references see eg. \cite{MARCELA}), which
are interesting to remember, particularly since some of them
are now more relevant because of the heaviness of the top quark.
First of all, there are quite general (and therefore rather
weak) upper bounds which follow from the requirement of a
unitary and weakly interacting theory at $M_Z$. We get then
$M_h\leq {\cal O}(1$ TeV). More interesting bounds are scale dependent
and are known under the name of triviality (upper) bound and
vacuum stability (lower) bound. They are shown in Fig.1 for
several values of the cut-off scale $\Lambda$ (the bounds are
obtained by requiring perturbative theory, with a stable vacuum,
up to the scale $\Lambda$). 

\begin{figure}
\center
\psfig{figure=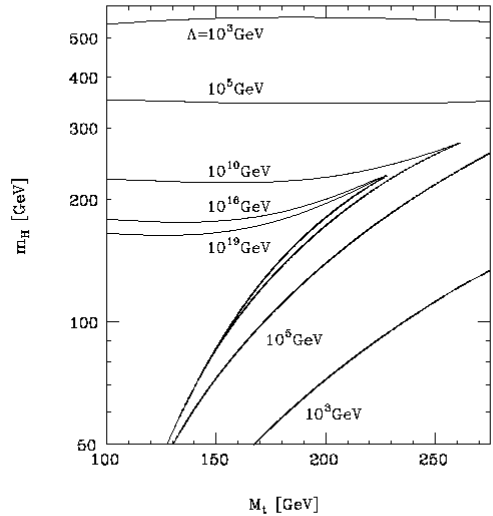,height=8.0cm}
\caption{Triviality (upper) and stability (lower) bounds on the SM Higgs boson
mass as a function of the top-quark mass for different cut-off scales 
$\Lambda$. The figure has been taken from ref. \protect\cite{MARCELA}.}
\label{fig:roch1}
\end{figure}

Particularly strong are the bounds
if the SM is to be valid up to the GUT scale: 
for $m_t=(175\pm6)$ GeV one gets 140 GeV $\leq
M_h\leq 180$ GeV. If $M_h$ was smaller than 140 GeV then this would
give us a direct information on the presence of new physics at
scales below $M_{GUT}$. For instance, as seen in Fig.1, $M_h
\cong 80$ GeV would imply new physics below 1 TeV (but, in general,
not vice versa; the supersymmetric Higgs sector is discussed
later). 

\subsection{Scattered clouds on the status of the SM}

The overall global succes of the SM is a bit overshadowed by a
couple of (quite relevant!) scattered clouds. The results for
$R_b$ are still preliminary and differ by about 2$\sigma$ between
different experiments. The world average is only 1.8$\sigma$ away
from the SM prediction but with the error which is the same as
the maximal possible enhancement of $R_b$ in the Minimal
Supersymmetric Standard Model (see later). 

The effective Weinberg angle is now reported with a  very high
precision. However, this result comes from averaging over the
SLD and LEP results which are more then 3$\sigma$ apart.
Moreover, a crucial role in the final result (and in particular,
in the smallnes of the error) is played by the measurement of
the forward-backward asymmetry in the $b\bar{b}$ channel.
However, the value of the so-called ${\cal A}_b$ parameter:
\begin{eqnarray}
{\cal A}_b={(g^b_L)^2-(g^b_R)^2\over (g^b_L)^2+(g^b_R)^2} 
\end{eqnarray}
where
\begin{eqnarray}
{\cal L}_{Zb\bar{b}}\equiv {e\over4s_Wc_W}Z^0_\mu
\bar{b}\gamma^\mu(g^b_L(1-\gamma_5)+g^b_R(1+\gamma_5))b\nonumber
\end{eqnarray}
extracted from the same measurement is $3\sigma$ below the 
SM prediction. One may have, therefore, some doubts whether
systematic errors in the measurement of $A^{b,0}_{FB}$ have been
properly taken into account (e.g. QCD corrections to this
observable may cause some problems). If this measurement is
omitted from the final average over $\sin^2\theta_{eff}^{lept}$,
then the result is closer to the SLD value and with larger
error. We shall return to the relevance of the result for 
$\sin^2\theta_{eff}^{lept}$ for new physics in the context of the MSSM.
 
Of course, it is not totally excluded that the present value 
of ${\cal A}_b$ is the correct one. This possibility looks 
unlikely, though. Indeed
\begin{eqnarray}
{\cal A}_b= {\cal A}^0_b + 2
{(1-{\cal A}^0_b)g^0_L\epsilon_L-(1+{\cal A}^0_b)g^0_R\epsilon_R
\over (g^0_L)^2+(g^0_R)^2}
\end{eqnarray}
where the superscript "$0$" denotes the SM value,
$g_{L,R}=g^0_{L,R}+\epsilon_{L,R}$ and physics beyond the SM
contributes to $\epsilon_{L,R}$. It is difficult to obtain large
negative $\delta {\cal A}_b$ since ${\cal A}^0_b\approx 1$ 
and $g^0_R\approx0$. Morover, 
\begin{eqnarray}
\delta {\cal A}_b\approx  5.84\times(1-{\cal A}^0_b)\times \delta R_b 
\end{eqnarray}
if $\epsilon_L\neq 0$, $\epsilon_R=0$, and
\begin{eqnarray}
\delta {\cal A}_b\approx -5.84\times(1+{\cal A}^0_b)\times \delta R_b 
\end{eqnarray}
if $\epsilon_R\neq 0$, $\epsilon_L=0$.
Thus, e.g. for $\delta R_b\approx 0.002$~, we can get 
$\delta {\cal A}_b\approx-0.023$ but 
we need a large (tree level ?) effect on $g_R$ and 
$\epsilon_L\sim 0$ \cite{CLINE}.

\section{Electroweak Interactions\\ and Physics Beyond the SM}

The overall success of the SM and, in particular, the new values of
$R_b$ and $R_c$ reported at this conference, have an important
impact on speculations on new physics. Any purely experimental
motivation for new physics has disappeared. Therefore, I am not
going to discuss {\it ad hoc} models suggested to explain previous
values of $R_b$ and $R_c$. Similarly, the direct phenomenological 
interest in  the models with $Z^{0\prime}$ has evaporated: the
generation-blind couplings are no longer supported and
leptophobic $Z^{0\prime}$ has become now both leptophobic and
hadrophobic (of course, there is some continous theoretical
interest in models with $Z^{0\prime}$ which remains unchanged; such
models need not imply any strong phenomenological consequences
at the electroweak scale). With the  experimental results no
longer {\it pressing} for new physics, the proper question to ask now
is what {\it room} do they leave for physics beyond the SM close to
the electroweak scale. This question is important for those extensions
of the SM which have independent theoretical motivation and, of
course, for future experiments at LEP2 and LHC. Since
technicolour ideas have already been discussed \cite{LANE}, in this talk
I focuss on supersymmetry.

\subsection{Introduction to the Minimal Supersymmetric\\ Standard Model}

Supersymmetry is of interest for a number of reasons. It is
likely that it is linked to the electroweak symmetry breaking
(hierarchy problem). It is at present the only theoretical framework
which allows to extrapolate to very short distances (Planck
length). It is an appealing mathematical structure.
Supersymmetric field theories have several interesting
properties which make them more predictive that
non-supersymmetric theories. And finally, on the purely pragmatic
level, the Minimal Supersymmetric Standard Model is so far the
only framework beyond the SM which addresses the phenomenology
of elementary interactions at the electroweak scale and just
above it in a complete and quantitative way. As such, it plays
an important stimulating role in experimental search for physics
beyond the SM.
 
The minimal model is based on the three main assumptions: 
\begin{itemize}
\item[a)] minimal particle content consistent with the known
spectrum and supersymmetry
\item[b)] most general soft supersymmetry breaking terms which
are $SU_c(3)\times SU_L(2)\times U_Y(1)$ symmetric
\item[c)] $R$-parity conservation
\end{itemize}
Two "mild" extensions of the minimal model include the models with
$R$-parity explicitly broken and those with additional full $SU(5)$
matter multiplets, at "low" scale. New $R$-parity violating
couplings in the superpotential must be small enough to avoid
problems with the baryon and lepton number violating processes.
Additional complete $SU(5)$ multiplets do not destroy the
unification of couplings.

\subsection{MSSM and precision data}

The simplest interpretation of the success of the SM is that the 
superpartners are heavy enough to decouple from the electroweak
observables. Explicit calculations (with the same precision as
in the SM) show that this happens if the common supersymmetry breaking 
scale is $\geq {\cal O}(300-400)$ GeV. This is very important
as such a scale of supersymmetry breaking is still low enough for
supersymmetry to cure the hierarchy problem. However, in
this case the only supersymmetric signature at the electroweak
scale and just above it is the Higgs sector. The Higgs sector in
the MSSM has been extensively studied in the recent years. The  
one- \cite{ONEL} and two-loop
\cite{TWOL} corrections to the Higgs boson masses have been
calculated and a light Higgs boson, $M_h\leq {\cal O}(150)$ GeV, 
is the most firm prediction of the MSSM. This prediction is
consistent with the SM fits discussed earlier. We can, therefore, 
conclude at this point that the supersymmetric extension 
of the SM, with all superpartners $\geq {\cal O}(300)$ GeV, is
phenomenologically as succesful as the SM itself and has the
virtue of solving the hierarchy problem. Discovery of a light
Higgs boson is the crucial test for such an extension.

The relatively heavy superpartners discussed in the previous
paragraph are sufficient for explaining the success of the SM.
But is it necessary that all of them are that heavy? Is there a
room for some light superpartners with masses ${\cal O}(M_Z)$ or 
even below?  This question is of great importance for LEP2. Indeed,
a closer look at the electroweak observables shows that the
answer to this question is positive. The dominant quantum
corrections to the electroweak observables are the so-called
"obligue" corrections to the gauge boson self-energies. They are
economically summarized in terms of the $S,T,U$  parameters
\begin{eqnarray}
S\sim \Pi^\prime_{3Y}(0)=\Pi^\prime_{L3,R3}
+\Pi^\prime_{L3,B-L}
\end{eqnarray}
(the last decomposition is labelled by the
$SU_L(2)\times SU_R(2)\times U_{B-L}(1)$ quantum numbers)
\begin{eqnarray}
\alpha T\equiv\Delta\rho\sim\Pi_{11}(0)-\Pi_{33}(0)
\end{eqnarray}
\begin{eqnarray}
U\sim \Pi^\prime_{11}(0)-\Pi^\prime_{33}(0)
\end{eqnarray}
where $\Pi_{ij}(0)$ $(\Pi^\prime_{ij}(0))$ are the (i,j) left
-handed gauge boson self-energies at the zero momentum (their
derivatives) and the self- energy correction to the $S$
parameter mixes $W^\pm_\mu$ and $B_\mu$ gauge bosons. It is clear
from their definitions that the parameters $S,T,U$ have important
symmetry properties: $T$ and $U$ vanish in the limit when quantum
corrections to the left-handed gauge boson self-energies have
unbroken "custodial" $SU_V(2)$ symmetry. The parameter $S$ vanishes
if $SU_L(2)$ is an exact symmetry (notice that, 
since ${\bf 3_L}\times {\bf 3_R}={\bf 1}\oplus {\bf 5}$ under 
$SU_V(2)$, exact $SU_V(2)$ is not sufficient for the vanishing
of $S$). The success of the SM means that it has just the right
amount of the $SU_V(2)$ breaking (and of the $SU_L(2)$ breaking),
encoded mainly in the top quark-bottom quark mass splitting. Any
extension of the SM, to be consistent with the precision data,
should not introduce additional sources of large $SU_V(2)$ breaking
in sectors which couple to the left-handed gauge bosons. In the
MSSM, the main potential origin of new $SU_V(2)$ breaking effects
in the left-handed sector is the splitting between the
left-handed stop and sbottom masses:
\begin{eqnarray}
M^2_{\tilde{t}_L}=m^2_Q+m^2_t-\cos2\beta (M^2_Z-4M^2_W)\nonumber\\ 
M^2_{\tilde{b}_L}=m^2_Q+ m^2_b-\cos2\beta (M^2_Z+2M^2_W)
\end{eqnarray}
The $SU_V(2)$ breaking is small if the common soft mass $m^2_Q$ is 
large enough. So, from the bulk of the precision data one gets a lower 
bound on the masses of the left-handed squarks of the third 
generation
\footnote{Additional source of the $SU_V(2)$ breaking is in
the $A$-terms. In principle, there can be cancellations between the 
soft mass terms and the $A$-terms, such that another solution with small
$SU_V(2)$ breaking exists with a large inverse hierarchy 
$m^2_U\gg m^2_Q$. This is very unnatural from the point of view of the
GUT boundary conditions and here we assume $m^2_Q>m^2_U$.}.
However, the
right-handed squarks can be very light, at their experimental
lower bound $\sim 45$ GeV. Another interesting observation is 
that in the low $\tan\beta$
region the top squark masses are strongly constrained also by the 
present experimental lower bound on the lightest supersymmetric Higgs
boson  mass, $M_h\geq 60$ GeV. For low $\tan\beta$, the tree level
Higgs boson mass is close to zero and radiative corrections are
very important. They depend logarithmically on the product
$M_{\tilde{t}_1}M_{\tilde{t}_2}$. 
\begin{figure}
\center
\psfig{figure=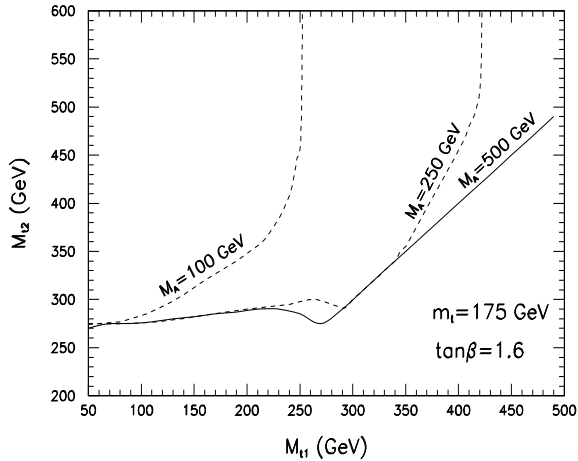,height=7.0cm}
\caption{Lower bounds on the heavier stop, $\tilde{t}_2$, as a function
of $M_{\tilde{t}_1}$. The solid line shows the bound from $M_h>60$ GeV and 
the requirement of a good $\chi^2$ fit to the electroweak observables. The 
dashed lines show the bounds obtained
from the $b\rightarrow s\gamma$ constraint for two values of the $CP$-odd
Higgs boson mass $M_A$. The latter bound is obtained under the assumption 
that the chargino masses $m_{C^\pm_1}=90$ GeV.}
\label{fig:roch2}
\end{figure}
In Fig.2 we show the lower bound  
on the mass of the heavier top squark as a function of the mass of 
the lighter stop,  which follows  from the
requirement that a fit in the MSSM is at most by $\Delta\chi^2 =2$
worse than the analogous fit in the SM and from the lower bound
on $M_h$. The limits on the stop masses obtained from the bound on 
$M_h$ are of similar strength as the $\chi^2$ limits but apply only 
for low  $\tan\beta$. The important role played in the fit by
the precise result for $\sin^2\theta_{eff}^{lept}$ is illustrated 
in Fig.3. The world average value (used in obtaining the bounds
shown in Fig.2) is obtained in the SM model with
$m_t=(175\pm6)$ GeV and $M_h\sim (120-150)$ GeV, with
little room for additional supersymmetric contribution. Hence,
the relevant superpartners ($\tilde{t}_L$ and $\tilde{b}_L$)
have to be heavy. With lighter superpartners, one obtains the
band (solid lines) shown in Fig.3. We see that the SLD result for
$\sin^2\theta_{eff}^{lept}$ leaves much more room for light
superpartners. Thus, settling the SLD/LEP dispute is very
relevant for new physics.

\begin{figure}
\center
\psfig{figure=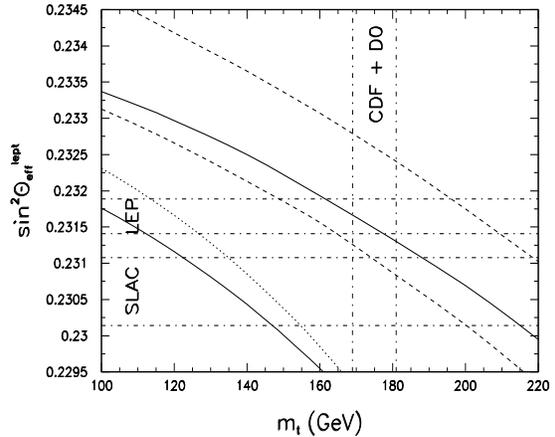,height=7.0cm}
\caption{Predictions for $\sin^2\theta^{lept}_{eff}$ in the SM 
(the band bounded by the dashed lines) and in the MSSM (solid lines) as 
functions of the top-quark mass.  Dotted line shows the lower limit in 
the MSSM if all sparticles are heavier than $Z^0$.
The SLC and the (average) LEP measurements are
marked by horizontal dash-dotted lines.}
\label{fig:roch3}
\end{figure}

All squarks of the first two generations as well as sleptons can
be very light, $\leq {\cal O}(M_Z)$, and the success of the SM 
in the description of the precision electroweak data is still
maintained. The same applies to the gaugino/higgsino sectors,
since they do not give any strong $SU_V(2)$ breaking effects. In 
conclusion, most of the superpartners decouple from most of the
electroweak observables, even if very light, $\leq {\cal O}(M_Z)$. 
This high degree of screening follows from the basic structure of 
the model.

The remarkable exception is the famous $R_b$ \cite{RB}. Additional
supersymmetric contributions to the $Z^0\bar{b}b$ vertex,
precisely from the chargino-right-handed stop
loop, can be non-negligible 
when both are light (and from the $CP$-odd Higgs loop in the
large $\tan\beta$ region). (Note that those contributions do not 
change the value of ${\cal A}_b$
as they dominantly modify the left-handed effective coupling.)
However, even with the chargino and stop very light, at their
present experimental mass limit, in the MSSM 
the prediction for $R_b$ depends on the chargino composition
and on the stop mixing angle. The values ranging from 0.2158 (the SM
prediction) up to 0.218 (0.219) for small (large) $\tan\beta$
can be realistically obtained (given all the experimental 
constraints) \cite{GRA}.
\begin{figure}
\center
\psfig{figure=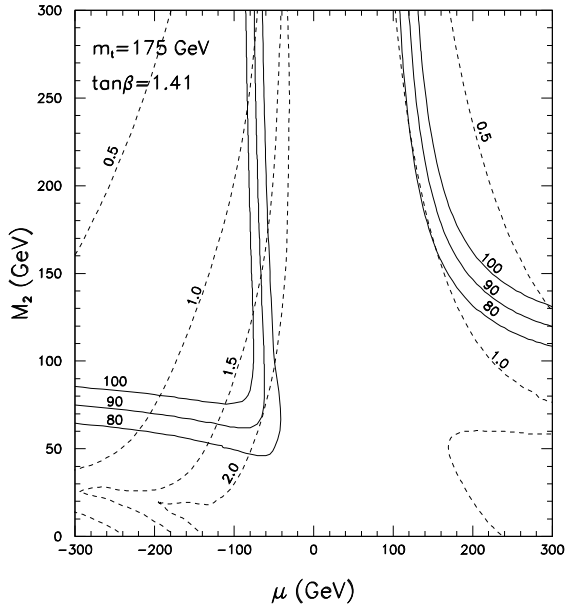,height=9.0cm}
\caption{Contours of constant lighter chargino masses
$m_{C^\pm_1}=$80, 90, 100 GeV (solid lines) and of
$\delta R_b\times10^3=$2.0, 1.5, 1.0, 0.5 
(dashed lines) in the $(\mu, M_2)$ plane for $m_t=175$ GeV 
and $\tan\beta=\sqrt2$. The region below the lines 
$m_{C^\pm_1}=$80 GeV 
is excluded after the LEP run at $\sqrt s=161$ GeV.}
\label{fig:roch4}
\end{figure}
No significant modification of the SM result for $R_c$ is possible, 
though. This predictions hold with or without
$R$-parity conservation and with or without the GUT relation for the 
gaugino masses. The upper bound is reachable for chargino masses
up to ${\cal O}(90$ GeV) provided they are mixed gaugino-higgsino states
$(M_2/|\mu|\sim1)$. In the same chargino mass range
$\delta R_b\rightarrow 0$ in the deep higgsino and gaugino
regions. Clearly, the new values of $R_b$ and $R_c$ are good
news for supersymmetry! At the same time, one should face the
fact that, unfortunately, in the MSSM
\begin{eqnarray}
\delta R_b^{max}\sim {\cal O}(1 ~\sigma^{exp})\nonumber
\end{eqnarray}
so much better experimental precision is needed for a meaningful
discussion. The contours of $\delta R_b$ in the $(M_2,\mu)$
plane are shown in Fig.4.

\subsection{Other effects of light superpartners?}

There are several well known supersymmetric contributions to
rare processes. In particular, in the soft terms supersymmetry 
may provide new sources of flavour violation. However, even
assuming the absence of such new effects, there are obvious new
contributions when the  $W^\pm - q$ SM loops are replaced by the
$H^\pm - q$ loops and by the $\tilde{W}^\pm(\tilde{H}^\pm)
-\tilde{q}$ loops. Those can be 
expected to be very important in the presence of a light chargino
and stop and they contribute to all best measured
observables: $\varepsilon_K$-parameter for the $\bar{K}^0$-$K^0$
system, $\Delta m_B$ from $\bar{B}^0$-$B^0$ mixing and
$BR(b\rightarrow s\gamma)$.

There are two important facts to be remembered about these
contributions. They are present even if quark and squark mass
matrices are diagonal in the same super-Kobayashi-Maskawa basis.
However,  the coupling in the vertex $d_i\tilde{u}_jC^-$ depends on
this assumption and can depart from the $K$-$M$ parametrization if
squark mass matrices have flavour-off diagonal entries in the
super-Kobayashi-Maskawa basis. Some of those entries are still
totally unconstrained and this is precisely the case for the
(right) up squark mass matrix, that is relevant e.g. for the
couplings $b\tilde{t}_RC^-$. Still, sizeable suppression
compared to the $K$-$M$ parametrization requires large flavour-off
diagonal mass terms, of the order of the diagonal ones. To
remain on the conservative side,  we discuss the effects of a light
chargino and stop on rare processes under the assumption of the
$K$-$M$ parametrization of the chargino vertices. The recent
completion of the next-to-leading order QCD corrections
\cite{MISIAK} to
BR$(b\rightarrow s\gamma )$ leaves much less room for new
physics which gives an increase in this branching ratio and
favours new physics which partially cancels the Standard Model
contribution to the $b\rightarrow s\gamma$ amplitude. This is
generic for chargino-stop contributions with light particles (for
low $\tan\beta$) whereas the charged Higgs contribution
adds positively to the SM contribution.

It is interesting to present the lower bounds on the $CP$-odd
scalar mass $M_A$ (which to a good approximation is in one to
one correspondence with the charged Higgs boson mass) for
several values of the chargino and lighter stop masses and as a
function of the chargino composition measured by the ratio
$r=M_2/|\mu|$. For $m_{C^+}=M_{\tilde{t}_1} >$ 500 GeV, the
chargino - stop loop contribution is totally suppressed and we
get the  strong bound $M_A\geq 500$ GeV. For a lighter stop and
chargino, their contribution can cancel positive contribution
from the charged Higgs exchange and the limit on $M_A$ is much
weaker, except for the gaugino - like charginos which contribute
weakly to $b\rightarrow s\gamma$ (see Fig.5).
\begin{figure}
\center
\psfig{figure=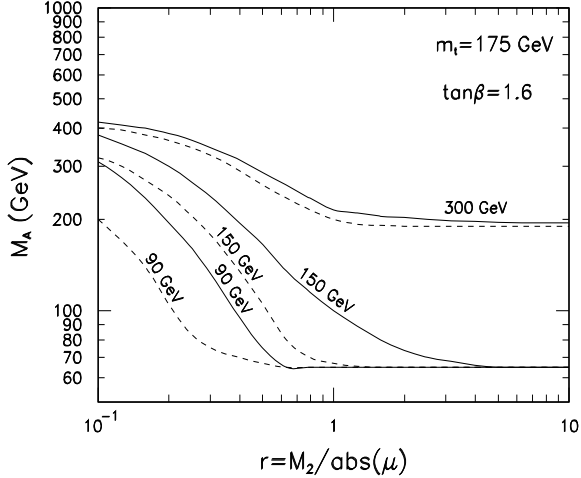,height=7.0cm}
\caption{Lower bounds on the $CP$-odd Higgs boson mass, $M_A$, as 
a function of the lighter chargino composition $r\equiv M_2/|\mu|$ 
for $m_{C^\pm_1}=M_{\tilde t_1}=$ 90, 150 and 300 GeV  for negative
(solid lines)  and positive (dashed) values of $\mu$.}
\label{fig:roch5}
\end{figure}

The second important remark is that the element $V_{td}\approx
A\lambda^3(\rho -i\eta)$ (in the Wolfenstein parametrization),
which is necessary for the calculation of the chargino and
charged Higgs boson loop contribution to the $\bar{\epsilon}_K$
parameter and the $\bar{B}^0$-$B^0$ mixing, is not directly
measured. Its SM value can change after the inclusion of new
contributions. Thus the correct approach is the following one:
take e.g.
\begin{eqnarray}
\Delta m_B\approx
f^2_{B_d}B_{Bd}\mid V_{tb}V^\star_{td}\mid^2\mid\Delta\mid
\end{eqnarray}
where
\begin{eqnarray}
\Delta =\Delta_W+\Delta_{NEW}
\end{eqnarray}
is the sum of all box diagram contributions, $f_{B_d}$ and
$B_{B_d}$ are the
$B^0$ meson decay constant and the vacuum saturation parameter.
The CP violating parameter $\varepsilon_K$ can also be expressed
in terms of $\Delta$. Given $\mid V_{cb}\mid$, and 
$|V_{ub}/V_{cb}|$ (known from the tree level processes i.e.
almost unaffected by the supersymmetric contributions) one can
fit the parameters $\rho ,\eta$ and $\Delta$ to the experimental
values of $\Delta m_{B_d}$ and $\mid\varepsilon_K\mid$. This way we
find \cite{BRANCO} a model independent constraint
\begin{eqnarray}
\frac{\Delta}{\Delta_W}< 3
\end{eqnarray}
for $\sqrt{f^2_{B_d}B_{B_d}}$ in the range (160 - 240) GeV and $B_K$
in the range (0.6 - 0.9) GeV.  In the next step, this result  can be used
to limit the allowed range of the stop and chargino masses and
mixings. The parameter space which is relevant for an increase
in $R_b$ gives large contribution to $\Delta$. It is still
consistent with the bound but requires modified (compared to
the SM) values of the $CP$-violating phase $\delta(\eta,\rho )$.

\subsection{Have light superpartners already been discovered?}

A considerable attention during this conference has been paid
to the two interesting pieces of experimental information: a
single event $e^+e^-\gamma\gamma + mising ~E_T$ has been reported
by the CDF and the results from the LEP 1.5 run at $\sqrt{s}=
136$ GeV include peculiar four jet events reported by ALEPH.
Both findings should be taken with extreme caution and are
likely to be a statistical fluctuation
\footnote{At the time of this
writing (October 96) no more two-photon events have been
reported. Four-jet events are neither ruled out nor
confirmed after the run at $\sqrt{s}=161$ GeV.}.
Nevertheless, they
generated some speculations on being a possible manifestation of
supersymmetry. The CDF event can be interpreted as a
selectron pair production with a subsequent chain decay:
\begin{eqnarray}
p\bar{p}\rightarrow\tilde{e}^+\tilde{e}^-\rightarrow
(e^+X_2)(e^-X_2)\rightarrow \nonumber\\
\rightarrow (e^+X_1\gamma )(e^-X_1\gamma )
\end{eqnarray}
where $X_1$ is the LSP which carries the missing energy and
$X_2$ is next-to-the LSP particle. The event can be
interpreted in two possible ways \cite{KANE}:
\begin{itemize}
\item[a)] $X_2$ - neutralino (gaugino)\\
$X_1$ - neutralino (higgsino)
\end{itemize}
The signatures of the event are reproduced for
$m_{X_2}-m_{X_1} \geq 30$ GeV, $\tan\beta\sim 1$ and
"nonunified" gaugino masses $M_1\approx M_2$. This
interpretation is consistent with a light supersymmetric
spectrum of the type discussed earlier (in particular, the one
which may give some enhancement in $R_b$).
\begin{itemize}
\item[b)] $X_2$ - neutralino\\
$X_1$ - gravitino ($\tilde G$) with BR$(X_2\rightarrow
{\tilde G}\gamma)\sim 1$,\\
$M_{X_2}\leq 100$ GeV, $m_{\tilde G}\leq 250$ eV (for 
$X_2$ to decay in the detector) 
\end{itemize}
This second interpretation fits nicely into the ideas of the
so-called gauge mediated low energy supersymmetry breaking 
\cite{DINE}.

The supersymmetric interpretation of the ALEPH four-jet events
is also possible, though not strikingly "natural". Their main
signatures (the absence of missing energy and of b-quark jets) can
be consistant with the so-called light gluino scenario
\cite{FARRAR} or otherwise needs broken $R$-parity.
In the latter case the events can be interpreted as a sneutrino pair 
production \cite{BARGER} or the right-handed stop pair production
(the cross section for the $\tilde{t}_R$ is a bit low but 
not necessarily inconsistent with the data averaged over all four experiments)
with subsequent $R$-parity violating decay into a pair of quarks,
or production of a pair of charginos which then decay
$C^+\rightarrow\tilde{t}_Rb\rightarrow qqb$, provided
$m_{C^+}\approx M_{{\tilde t}_R}$ so that the b-quark is slow
enough to be invisible \cite{MY}.
It is fair to wait for experimental clarification
before speculating further.

\section{Summary}
The $SM$ is impressively succesful in its global
description of the precision electroweak data. Nevertheless,
there still persist several important experimental uncertainties
in the observables which are very relevant for the theory,
in particular for new physics. These are $R_b$,
$\sin^2\theta_{eff}^{lept}$ and ${\cal A}_b$.

Fits to precision data give some indication for a light,
${\cal O}(100)$ GeV, Higgs boson but this evidence is much less
significant than the analogous prediction for the top quark
mass. The discovery of the Higgs boson and, particularly, its
mass remain important clues to physics beyond the SM. Light
versus heavy Higgs boson has its correspondence in supersymmetry
versus dynamical electroweak symmetry breaking.

Supersymmetric extension of the SM is not only theoretically
motivated but naturally accommodates the success of the SM, even
with some superpartners with masses $m\leq M_Z$. This is a
consequence of the structure of the theory and not of
fine-tuning of its parameters.

Very light superpartners (e.g. $C^\pm$, $\tilde{t}_R$, $N^0$, $\dots$)
may have important effects on few selected observables such as
$R_b$, $b\rightarrow s\gamma$ ,$B^0$-$\bar{B^0},\dots$ which, however,
require still better experimental accuracy to be confirmed.

We have a couple of exotic experimental observations which could
find their interpretation in the supersymmetric framework.
Extreme caution with any firm conclusion is, however, adviced
before further experimental clarification.

And finally, the supersymmetric extension of the SM provides at
present the only theoretical scheme which gives us consistent
and quantitative weak scale- GUT scale connection. Therefore,
such ideas as gauge coupling unification and infrared fixed
point structures can and have been extensively studied \cite{SP}, 
with clear connection to the low energy phenomenology. Moreover, 
the two pressing problems in supersymmetric phenomenology: the 
"theory" of soft terms and the flavour problem, which are likely 
to have their solution in physics at the high scale, can hopefully 
be studied experimentally at low energies!

\section*{Acknowledgments}

\noindent I am greatful to P.H. Chankowski  for fruitful collaboration and  
          his help in the preparation of this report. This work was partially 
          supported by the Polish Committee for Scientific Research.

\section*{Questions}

\noindent{\it Z. Lalak, Warsaw University:}\\
Can you comment on the allowed range of values of "$\tan\beta$"
in supersymmetric models ?

\vskip 12pt

\noindent{\it S. Pokorski:}

There are several constraints on very low ($\sim 1$) and very large ($>
m_t/m_b$) values of $\tan\beta$. For instance, with  $m_t=(175\pm 6)$
GeV, low values of $\tan\beta$ are excluded if we require
perturbative Yukawa coupling up to the GUT scale. Precise value
of the bound depends on the running of the strong coupling
constant. E.g.  new low mass $SU(5)$ multiplets push the bound
down, closer to one.

\vskip 12pt

\noindent{\it J.M. Izen, University of Texas at Dallas:}

Both you and Alain Blondel emphasized the importance of
measuring $\sigma(e^+e^-\rightarrow hadrons)$, especially in the
low energy region. In Beijing we are currently commissioning the
BES detector upgrade, and we will be running in 1996-97 on the
$J/\psi$ where we have the highest event rate. However, we plan to
take 2 weeks of the run to test the feasibility of an 
$R\equiv\sigma(e^+e^-\rightarrow hadrons)/\sigma(e^+e^-\rightarrow\mu^+\mu^-)$ 
scan with BES. In particular we study whether we can reduce the
energy requirement in our trigger. We have been using an energy
threshold of $\sim 1$ GeV for $J/\psi,\psi^\prime$ and 4.03
$D_s$, but this would be too high for a total cross section
measurement. We expect $\sim 1000$ events per day during an $R$
scan. It will be important to have a better continuum Monte
Carlo generator in this energy region for this measurement.

\section*{References}
\begin{thebibliography}{99}
\bibitem{BLON} A. Blondel, plenary talk, these Proceedings

\bibitem{HOL} W. Hollik, parallel session, these Proceedings\\
              B. Kniehl, parallel session, these Proceedings;\\
              both include an extensive list of references

\bibitem{SEE} See e.g. Proceedings of the 10th International Symposium 
              on High Energy Spin Physics, Nagoya, Japan 1992; Proceedings  
              of the International Symposium on the Future of Muon Physics;\\
              Heidelberg (1991), edited by K. Jungmann,\\
              V.W. Hughes and D. zu Putliz [J.Phys. {\bf C56} (1991)].

\bibitem{CZAR} A. Czarnecki, B. Krause, W. Marciano, 
               Phys.Rev. {\bf D52} (1995), 2619;\\
               S. Peris, M. Perrottet, E. de Rafael, Phys.Lett. {\bf B355}
               (1995), 523

\bibitem{EID} S. Eidelman, F. Jegerlehner, Z.Phys. {\bf C67} (1995), 585

\bibitem{HAY} M. Hayakawa, T. Kinoshita, A.I. Sanda, Phys.Rev.Lett. {\bf 75}
              (1995), 790,\\
              J. Bijnens, E. Pallante, J. Prades {\sl Phys. Rev. Lett.} 
              {\bf 75} (1995) 1447, {\bf E.} 3781, {\sl Nucl. Phys.} {\bf B474}
              (1996) 379.

\bibitem{NATH} P. Nath, parallel session, these Proceedings,\\
               M. Krawczyk, J. \.Zochowski preprint IFT-15/96 (hep-ph/9608321).

\bibitem{MARCELA} M. Carena et al., {\sl Higgs physics at LEP2}, in CERN yellow
                  report, CERN-96-01 (hep-ph/9602250).

\bibitem{MAN} M. Mangano, parallel session, these Proceedings

\bibitem{LANE} K. Lane, plenary talk, these Proceedings

\bibitem{ONEL} T. Okada, H. Yamaguchi, T. Yanagida 
                {\sl Prog. Theor. Phys. Lett.} {\bf 85} (1991) 1,\\
                H.E. Haber, R. Hempfling {\sl Phys. Rev. Lett.} {\bf 66} 
                (1991) 1815,\\
                J. Ellis, G. Ridolfi, F. Zwirner {\sl Phys. Lett.} 
                {\bf 257B} (1991) 83, {\bf 262B} (1991) 477,\\
                R. Barbieri, M. Frigeni, F. Caravaglios {\sl Phys. Lett.}
                {\bf 258B} (1991) 167,\\
                P.H. Chankowski, S. Pokorski, J. Rosiek
                {\sl Phys. Lett.} {\bf 274B} (1992) 191.

\bibitem{TWOL} R. Hempfling, A Hoang {\sl Phys. Lett.} {\bf 331B} (1994) 99,\\
               J. Kodaira, Y. Yasui, K. Sasaki {\sl Phys. Rev.} 
               {\bf D50} (1994) 7035,\\
               J.A. Casas, J.R. Espinosa, M. Quiros, A. Riotto
               {\sl Nucl. Phys.} {\bf B436} (1995) 3,\\
               M. Carena, J.R. Espinosa, M. Quiros, C.E.M. Wagner 
               {\sl Phys. Lett.} {\bf 355B} (1995) 209,\\
               M. Carena, M. Quiros, C.E.M. Wagner 
               {\sl Nucl. Phys.} {\bf B461} (1996) 407.

\bibitem{RB} M. Boulware, D. Finnell {\sl Phys. Rev.} {\bf D44} (1991) 2054,\\
             J. Rosiek {\sl Phys. Lett.} {\bf 252B} (1990) 135,\\
             A. Denner et al. {\sl Z. Phys.} {\bf C51} (1991) 695,\\
               G.L. Kane C. Kolda J.D. Wells.
               {\sl Phys. Lett.} {\bf338B } (1994) 219,\\
               P.H. Chankowski, S. Pokorski in Proceedings of the 
               Beyond the Standard Model IV conference, 
               Lake Tahoe C.A., eds. J.F. Gunion, T. Han, J. Ohnemus
               (1994) p. 233,\\
               D. Garcia, R. Jimen\'ez J. Sol\`a {\sl Phys. Lett.} {\bf347} 
               (1995) 309, 321, E {\bf351B} (1995) 602,\\
               D. Garcia, J. Sol\`a {\sl Phys. Lett.} {\bf354B} (1995) 335,
               {\bf357B} (1995) 349,\\
               G.L. Kane R.G. Stuart, J.D. Wells
               {\sl Phys. Lett.} {\bf 354B} (1995) 350. 
               P.H. Chankowski, S. Pokorski {\sl Phys. Lett.} {\bf 366B} 
               (1996) 188,\\
               G.L. Kane, J.D. Wells {\sl Phys. Rev. Lett.} {\bf 76}
               (1996) 869,\\
               J. Ellis, J.L. Lopez, D.V. Nanopoulos {\sl Phys. Lett.} 
               {\bf 372B} (1996) 95.

\bibitem{GRA} P.H. Chankowski, S. Pokorski {\sl Nucl. Phys.} {\bf B475}
              (1996) 3.

\bibitem{CLINE} J. Cline, parallel session, these Proceedings, \\
                P. Bamert et al. {\sl Phys. Rev.} {\bf D54} (1996) 4275.

\bibitem{MISIAK} M. Misiak, parallel session, these Proceedings,\\
                 K.G. Chetyrkin, M. Misiak, M. M\"unz to be published,\\
                 C. Greub, T. Hurth, D. Wyler {\sl Phys. Rev.} {\bf D54}
                 (1996) 3350,\\
                 C. Greub, T. Hurth talk given at 1996  Meeting of the 
                 Division of Particles and Fields of the American Physical 
                 Society, Minneapolis, August 1996 (hep-ph/9608449). 
 
\bibitem{BRANCO} G.C. Branco, G.C. Cho, Y. Kizukuri, N. Oshimo
               {\sl Phys. Lett.} {\bf 337B} (1994) 316, 
               {\sl Nucl. Phys.} {\bf B449} (1995) 483,\\
               G.C. Branco, W. Grimus, L. Lavoura {\sl Phys. Lett.} 
               {\bf 380} (1996) 119,\\
               A. Brignole, F. Feruglio, F. Zwirner {\sl Z.Phys.}
               {\bf C71} (1996) 679,\\
               J. Rosiek, parallel session, these Proceedings

\bibitem{KANE} S. Ambrosanio et al. {\sl Phys. Rev .Lett.} {\bf 76} (1996) 
               3498, {\sl Phys.Rev.} {\bf D54} (1996) 5395
               and  hep-ph/9607414, \\
               S. Dimopoulos, M. Dine, S. Raby, S. Thomas 
               {\sl Phys.Rev.Lett.} {\bf 76} (1996) 3494,\\
               S. Dimopoulos, S. Thomas, J. D. Wells {\sl Phys. Rev.} 
               {\bf D54} (1996) 3283,\\
               G.Kane, parallel session, these Proceedings

\bibitem{DINE} M. Dine, W. Fischler, M. Srednicki {\it Nucl. Phys.}
               {\bf B189} (1981) 575,\\
               S. Dimopoulos, S. Raby {\it Nucl. Phys.} {\bf B192}
               (1981) 353,\\
               L. Alvarez-Gaum\'e, M. Claudson, M. Wise {\it Nucl. Phys.}
               {\bf B207} (1982) 96,\\
               M. Dine, A. Nelson {\sl Phys. Rev.} {\bf D48} (1993) 1277,\\
               M. Dine, A. Nelson, Y. Shirman {\sl Phys. Rev.} {\bf D51}
               (1995) 1362,\\
               M. Dine, A. Nelson, Y. Nir, Y. Shirman {\sl Phys. Rev.}
               {\bf D53} (1996) 2658.

\bibitem{FARRAR} G.R. Farrar preprint RU-95-82 (hep-ph/9512306).

\bibitem{BARGER} V. Barger, W.-Y. Keung, R.J.N. Phillips {\sl Phys. Lett.}
                 {\bf 364B} (1995) 27.

\bibitem{MY} D.K. Ghosh, R.M. Godbole, S. Raychaudhuri, hep-ph/9605460,\\
             P. Chankowski, D. Choudhury, S. Pokorski preprint
             MPI-PTh/96-44, SCIPP-96/27 (hep-ph/9606415).
             H. Dreiner, S. Lola, P. Morawitz  hep-ph/9606364.

\bibitem{SP} S. Pokorski, talk given at International Workshop on Supersymmetry
              and Unification of Fundamental Interactions (SUSY 95), Palaiseau,
              France, May 1995 (hep-ph/9510224), \\
              G.G. Ross, plenary talk, these Proceedings
\end{thebibliography}
\end{document}